\documentclass[authoryear, review]{elsarticle}

\usepackage{amssymb}
\usepackage{amsmath}
\usepackage{multirow,subfigure, amsfonts,bbding}

\journal{Expert Systems with Applications}

\begin{document}

\begin{frontmatter}



\title{Target Speaker Lipreading by Audio-Visual Self-Distillation Pretraining
and Speaker
Adaptation}
%\tnotetext[label1]{

\author[snnu]{Jing-Xuan Zhang} %% Author name
\ead{jxzhanggg@snnu.edu.cn}

\author[ifly]{Tingzhi Mao} %% Author name
\ead{tzmao@iflytek.com}

\author[snnu]{Longjiang Guo} %% Author name
\ead{longjiangguo@snnu.edu.cn}

\author[snnu]{Jin Li} %% Author name
\ead{jin.li@snnu.edu.cn}

\author[snnu]{Lichen Zhang\corref{cor1}} %% Author name
\ead{zhanglichen@snnu.edu.cn}

\cortext[cor1]{Corresponding author.}

\affiliation[snnu]{organization={School of Computer Science, Shaanxi Normal University},%Department and Organization
            addressline={No. 620, West Chang'an Avenue, Chang'an District}, 
            city={Xi'an},
            postcode={710119}, 
            state={Shaanxi},
            country={P. R. China}}

\affiliation[ifly]{organization={iFLYTEK Research, iFLYTEK Co. Ltd.},%Department and Organization
            addressline={No. 666, Wangjiang West Road, Gaoxin District}, 
            city={Hefei},
            postcode={230088}, 
            state={Anhui},
            country={P. R. China}}
%% Abstract
\begin{abstract}

% Lipreading has wide applications such as facilitating human-computer interaction and communication in noisy environment.
% technology lends itself to facilitating natural human-computer interaction in home chat scenarios for smart devices or service robots.
% Lipreading technology tailored for home chat scenarios holds pivotal potential in enabling natural human-computer interaction for smart home devices and intelligent service robots
% Lipreading for home chat scenarios can be used for realizing natural human-computer interaction for smart home devices or intelligent service robots.% This study addresses the challenges associated with lipreading for target speakers in home chat scenarios, particularly in the Chinese language. 
% However, 
 % in the realistic environment, low-quality of video and variations in illumination, head poses as well as speech styles between different speakers
 % pose great challenge to this task.
% Lipreading recognition in realistic environment is challenging due to 
% the low-quality of video recordings from far-field cameras,  variations in illumination and speaker head poses, and colloquial speech styles. 

 Lipreading is an important technique for facilitating human-computer interaction in noisy environments. 
Our previously developed self-supervised learning method, AV2vec, which leverages multimodal self-distillation, has demonstrated promising performance in speaker-independent lipreading on the English LRS3 dataset. However, AV2vec faces challenges such as high training costs and a potential scarcity of audio-visual data for lipreading in languages other than English, such as Chinese.
Additionally, most studies concentrate on speaker-independent lipreading models, 
which struggle to account for the substantial variation in speaking styles across different speakers.
To address these issues, we propose a comprehensive approach. First, we investigate cross-lingual transfer learning, adapting a pre-trained AV2vec model from a source language and optimizing it for the  lipreading task in a target language. Second, we enhance the accuracy of lipreading for specific target speakers through a speaker adaptation strategy, which is not extensively explored in previous research. Third, after analyzing the complementary performance of lipreading with lip region-of-interest (ROI) and face inputs, we introduce a model ensembling strategy that integrates both, significantly boosting model performance. 
Our method achieved a character error rate (CER) of 77.3\% on the evaluation set of the ChatCLR dataset, which is lower than the top result from the 2024 Chat-scenario Chinese Lipreading Challenge.
% establishing a new benchmark for future research.
%Our method achieved a character error rate (CER) of 77.3\% on the ChatCLR dataset's evaluation set, lower than the top result from the 2024 Chat-scenario Chinese Lipreading Challenge and setting a new benchmark for future research.
%setting a new benchmark for future research in the field.

\end{abstract}

%Graphica%l abstract
\begin{graphicalabstract}
\includegraphics[width=\textwidth]{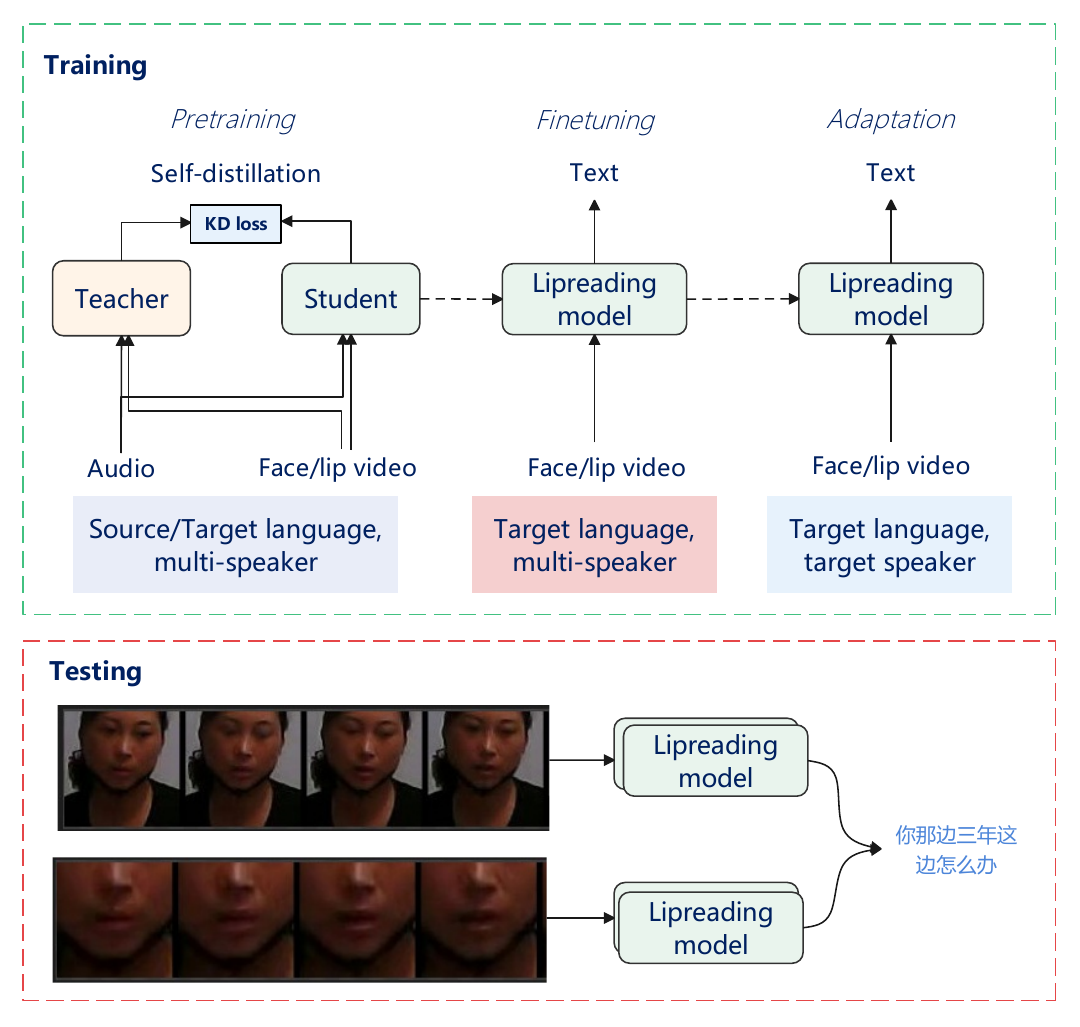}
\end{graphicalabstract}

\begin{highlights}
\item  Cross-lingual transfer learning enhances lipreading with limited target language data
\item Speaker adaptation  boosts specific speaker lipreading accuracy
\item Model ensembling with lip and face inputs improves lipreading
\item Our approach sets a new benchmark on the ChatCLR dataset
\end{highlights}

%% Keywords

\end{frontmatter}


\section{Introduction}
\label{sec:intro}
Lipreading, also referred to as visual speech recognition (VSR), aims at predicting spoken content from video of a speaker's lip movement 
without relying on any audio input~\citep{noda2014lipreading,wand2016lipreading,martinez2020lipreading,Themos2017combine,Zhang_Richmond_Ling_Dai_2021}. 
This task represents an important
approach towards developing a
silent speech interface (SSI)~\citep{denby2010silent}, which enables speech communication without
the need for vocalized sound production.
Since video is invariant to acoustic interference, lipreading has wide-ranging practical applications, such as facilitating human-computer communication and interaction with smart devices or intelligent service robots in noisy environments.
%, even in noisy environments.
% and communication with intelligent service robots, even in noisy environments.
%and enhancing the capabilities of hearing aids.

% Lipreading can be categorized into two main approaches: word-level~\cite{noda2014lipreading,  martinez2020lipreading,ma2022training} and sentence-level~\cite{Chung_2017_CVPR, zhang2019understanding, zhao2020hearing}.
% Word-level lipreading focuses on recognizing isolate words from a limited vocabulary, while  the sentence-level lipreading deals with continuous speech utterances with unconstrained vocabularies. 
% This work concentrates on the task 2 of 2024 Chat-scenario Chinese Lipreading (ChatCLR) Challenge\footnote{https://mispchallenge.github.io/ICME2024/}, which is a sentence-level lipreading task involving the target speaker. 
% The dataset was recorded in realistic home TV scenarios,
% capturing colloquial conversation in Chinese.
% This task poses several challenges. First, lipreading is inherently difficult due to the ambiguity of lip movements in discriminating between different words.
% For instance, words like "bark" and "park" are acoustically distinct but visually similar due to the many-to-one mapping from visemes to phonemes. 
% Second, videos were recorded by a far-field camera in a realistic home TV setting,
% resulting in substantial variations in speakers' head poses, illumination conditions, and emotional expressions.
% Third, the speakers' manner of speaking is highly casual and colloquial, deviating from formal speech patterns.

Recently, audio-visual self-supervised learning (SSL) has been proposed and applied to the lipreading task~\citep{ma2021lira, shiavhubert, zhu2023vatlm, hsu2022u}. 
This approach consists of a pretraining stage for extracting visual representations from unlabeled data, followed by a finetuning stage for lipreading task using labeled data. 
% A pre-text task is employed in
% pretraining stage such as contrastive learning~\cite{korbar2018cooperative} and masked prediction~\cite{shiavhubert}.
During the pretraining stage, self-supervised pretext tasks are employed, such as contrastive learning~\citep{korbar2018cooperative} and masked prediction~\citep{shiavhubert}, to facilitate effective representation learning.
While the initial motivation was to alleviate the issue of missing labels~\citep{baevski2020wav2vec,hsu2021hubert}, self-supervised learning has demonstrated its effectiveness in providing good initialization of the lipreading model parameters, even when all data is labeled~\citep{shiavhubert,Zhang2023av2vec}.
In our previous work, we proposed AV2vec~\citep{Zhang2023av2vec}, a self-supervised learning method for learning audio-visual representations by multimodal self-distillation. Our experimental results showed that the AV2vec model achieved better performance than the AV-HuBERT baseline on the LRS3 dataset~\citep{Afouras18d}. 
Regarding the inputs
for lipreading models, most works adopt lip region-of-interest (ROI)~\citep{afouras2018deep, ma2021end}. However, Zhang et al.'s work~\citep{zhangbeyondlip} suggests lipreading can benefit from additional clues within extraoral regions. Our previous work~\citep{Zhang2022face} also indicates that the model taking face video as input can achieve a better performance than those 
relying on traditional lip ROI under the self-supervised learning framework.

\begin{figure*}[t]
    \centering
    \includegraphics[width=0.9\textwidth]{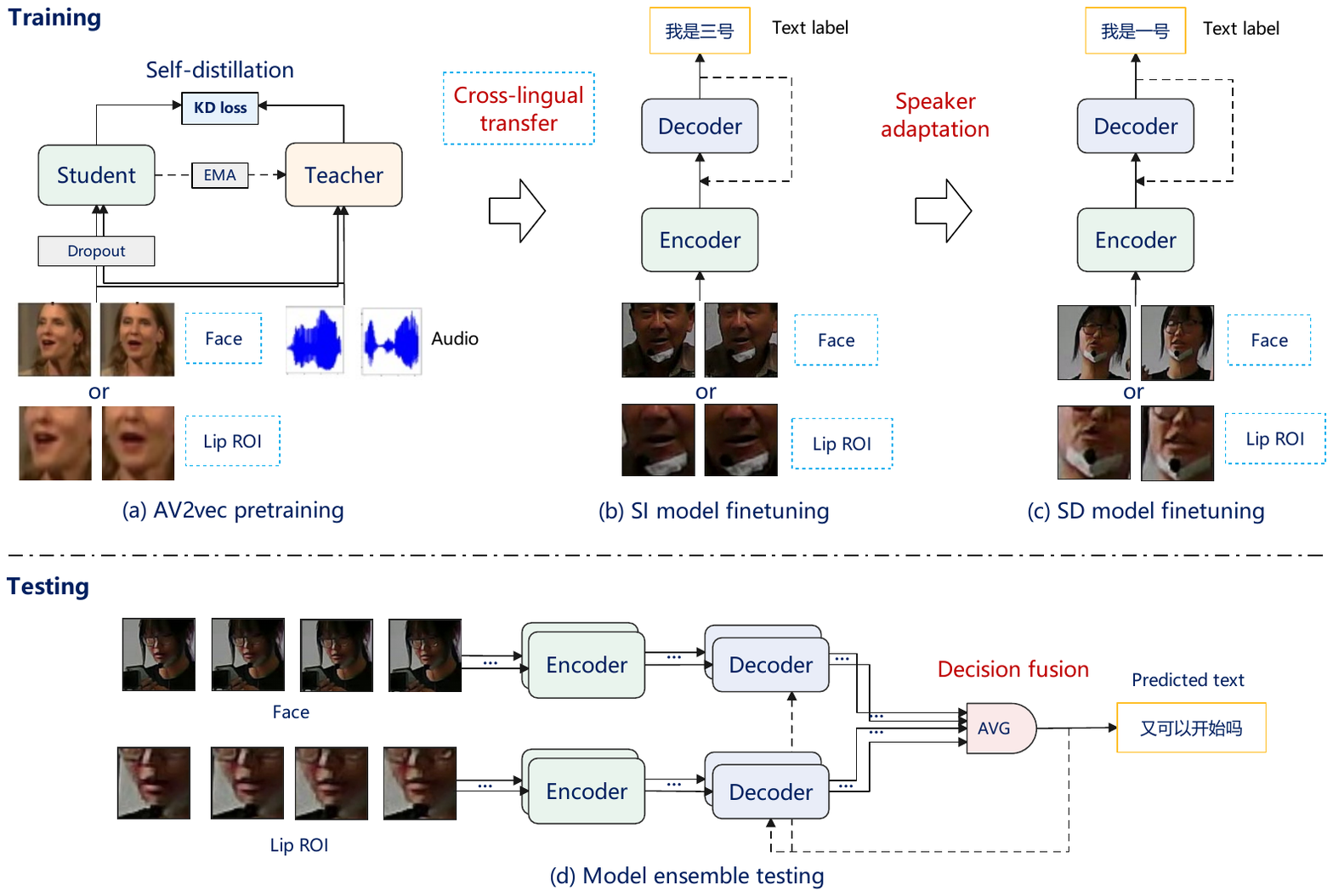}
    \caption{Illustration of our proposed method during training and testing. The dashed blue box indicates the optional configuration during model training. The dashed lines in (b), (c), and (d) illustrate the recurrent connections  within the auto-regressive decoder. The encoder is initialized using the pretrained student module, while the decoder is randomly initialized.
    Our model is trained with either the face or the lip region-of-interest video input. ``SI" and ``SD'' represent speaker-independent and speaker-dependent respectively.}
    \label{fig:figure0}
\end{figure*}

Despite the success of the previous studies on lipreading, several research gaps remain to be addressed.
First,
visemes, analogous to phonemes in speech, exhibit a high degree of cross-linguistic sharedness and commonality across different languages.
Therefore, this motivates us to
explore
transferring the knowledge learned from a pretrained model in a source language to a target language. However, this cross-lingual transfer learning strategy 
has not been well-explored for self-supervised learning-based lipreading approach.
Second, most previous lipreading studies have focused on the performance of speaker-independent model,
while few have investigated speaker adaptation of the model for  specific target speakers.
Speaker adaptation allows the model to be fine-tuned and tailored to the unique characteristics of a particular speaker, potentially improving performance and robustness.
Third, our previous studies have highlighted the advantage of utilizing 
full face videos as input for lipreading models,
rather than relying on the conventional lip ROI.
Nevertheless, face-based input may still suffer from the risk of introducing redundancy and interference comes from extra-lip regions. 

In this work, we tackle these research gaps from the following aspects.
First, we propose a cross-lingual transfer learning method for the AV2vec model, where an AV2vec model is
pretrained on the English LRS3 dataset, and subsequently transferred to the Chinese lipreading task, as shown in Figure~\ref{fig:figure0} (a) and (b).
We construct and compare two separate model variants: one employing full face input and the other utilizing the conventional lip region-of-interest (ROI) as the visual input.
Second, a speaker adaptation technique is employed for lipreading,
in which a speaker-independent (SI) model is finetuned on 
 the video data from the target speaker, resulting in speaker-dependent (SD) lipreading models as shown in Figure~\ref{fig:figure0} (c).
During adaptation, the Kullback-Leibler divergence (KLD)
between the speaker-independent and speaker-dependent
model's predictions is minimized as a regularization 
to prevent overfitting to the target speaker's data.
Third, we leverage an ensemble learning approach to synergistically combine lip region-of-interest (ROI) and face inputs, capitalizing on the complementary strengths of these two types of input to further enhance the robustness of lipreading, as shown in Figure~\ref{fig:figure0} (d). 
%Our proposed method is illustrated in Figure 1.

Our experiments validated the effectiveness of cross-lingual transfer learning for the self-supervised AV2vec model. While the transferred model did not achieve the performance of a model directly pretrained on a large amount of target language data, it significantly outperformed a model trained solely on limited target language data. This underscores the utility of cross-lingual transfer learning when target language data is scarce. Additionally, target speaker adaptation effectively enhanced the model's performance on the target speaker. Furthermore, model ensemble strategies for models with different input types, namely the full face and lip ROI video, significantly improved lipreading accuracy. 
On the 2024 ChatCLR Challenge evaluation set, 
our method achieved a character error rate (CER) of 77.3\%, which is lower than the top team in the challenge.
% Due to time constraints, we were unable to complete the training of our full model set. Nevertheless, our uncompleted method attained 4th place in the 2024 ChatCLR Challenge. 
%Eventually, our method achieved a character error rate of 77.3\% on the evaluation set of the 2024 ChatCLR Challenge, 
%surpassing the best result from the challenge.

The structure of this paper is as follows: Section 2 provides a review of the related works. Section 3 describes our previously proposed AV2vec method and our proposed method in this work is introduced in Section 4.
Experiments and results are detailed in Section 5. The limitation of our work is discussed in Section 6, followed by the conclusion in Section 7.

\section{Related works}
\subsection{Lipreading approaches with supervised learning}

In contrast to the pretraining and finetuning paradigm employed in self-supervised learning, lipreading approaches adopting supervised learning optimize models directly using labeled data.
% In contrast to the pretraining and finetuning paradigm of self-supervised learning, lipreading approaches with supervised learning optimized the model using labels directly.
Early approaches to lipreading adopted a two-stage paradigm involving feature extraction and classification~\citep{dupont2000audio, lee2004design, sterpu2018towards}. 
 With the advent of deep learning methods, these two stages have been fused into a single neural network-based model.
 Recent advances in lipreading technology have demonstrated remarkable progress in recognizing isolated words and short commands. Notable examples include 	
2DCNN+BiLSTM+ResNet+MLF~\citep{s23042284}, SyncVSR~\citep{ahn2024syncvsr}, and AVCRFormer~\citep{RYUMIN2024124159}.
These approaches have achieved state-of-the-art performance on the LRW dataset and have established fundamental techniques for audio-visual feature extraction and fusion that influence current research.
 While our approach shares the basic principle of visual feature extraction, we differ in focusing on lipreading of sentence-level continuous speech. Furthermore, word-level lipreading models have proven instrumental in initializing sentence-level lipreading systems.
%which have achieved state-of-the-art performance on the LRW dataset. Furthermore, word-level lipreading models have proven instrumental in initializing sentence-level lipreading systems.
%Lipreading for word or short command has been extensively studied~\citep{RYUMIN2024124159,s23042284,ahn2024syncvsr}, such as SyncVSR~\citep{ahn2024syncvsr} and AVCRFormer~\citep{RYUMIN2024124159}, which have achieved state-of-the-art performance on LRW corpus. 
%The word-level lipreading
%has also been broadly used for the initialization of sentence-level lipreading models. 
Currently, sentence-level lipreading is typically tackled under an end-to-end framework using sequence-to-sequence models, such as attention-based encoder-decoder (AED)~\citep{Chung_2017_CVPR, afouras2018deep,petridis2018end, xu2020discriminative}, connectionist temporal classification (CTC)~\citep{petridis2018audio, 8373881}, and recurrent neural network transducer (RNNT)~\citep{makino2019recurrent}. 
 These models consume raw pixel inputs from lip videos and produce text sequence outputs. Researchers have been endeavored to employ or design sophisticated neural network architectures for the lipreading task, such as Transformer~\citep{afouras2018deep, SerdyukBS22}, Conformer~\citep{ma2021end}, Branchformer~\citep{peng2022branchformer, wang2024mlca}, or visual Transformer pooling~\citep{prajwal2022sub}. Additionally,
% proposed to adopt Transformer or Conformer for lipreading. 
various training strategies have been proposed, including curriculum learning~\citep{afouras2018deep}, data augmentation~\citep{ma2021end, ma2022training, liu2023synthvsr}, knowledge distillation~\citep{afouras2020asr,ma2021towards}, using auxiliary visual targets~\citep{sterpu2020teach}, or predicting hidden representations from audio~\citep{ma2022visual}.

\subsection{Speaker adaptation for speech recognition}
Speaker adaptation for  speech recognition aims to enhance 
the recognition accuracy for specific speakers, which is  extensively explored 
in previous studies~\citep{bell2020adaptation}.
Various algorithms have been proposed for adapting neural network-based speech recognition models, including embedding-based~\citep{sari2020unsupervised}, model-based~\citep{swietojanski2016learning}, and data augmentation~\citep{huang2020rapid} approaches.
The model-based approach involves 
 optimization of model parameters to adapt to data from the target speaker, guided by the primary objective function. 
To  mitigate overfitting to the adaptation data, techniques such as LHUC~\citep{swietojanski2016learning} and  FHL~\citep{samarakoon2016factorized}
 have been introduced to reduce the number of speaker-dependent parameters. 
 Additionally, L2 loss~\citep{liao2013speaker} 
and KL divergence (KLD) loss~\citep{yu2013kl} 
are proposed to penalize deviations from the speaker-independent parameters. 
KLD regularization and multi-task learning methods have been employed for speaker adaptation in CTC and AED models~\citep{li2018speaker,meng2019}.
However, lipreading researches predominantly focus on speaker-independent models,  with limited validation of speaker adaptation techniques in the lipreading task.
A pioneering study by David et al.~\citep{gimeno2023comparing} investigates the use of Adapters for speaker adaptation in Spanish lipreading.
%One of poineer work is conducted by David et al.~\cite{gimeno2023comparing}, where the Adapters was used for speaker adaptation for VSR of Spanish.

\subsection{Audio-visual self-supervised learning}
Most audio-visual self-supervised learning methods leverage the natural synergy between the audio and visual channels of the video for cross-modal self-supervision.
The AVTS model proposes to learn general audio-visual representations via self-supervised temporal synchronization and a contrastive
loss~\citep{korbar2018cooperative}. Ma et al. propose a contrastive learning method for
both global and local audio-visual representations~\citep{NEURIPS2021_38ef4b66}. 
AV-HuBERT~\citep{shiavhubert}, an extension of  HuBERT~\citep{hsu2021hubert} designed for audio-visual speech data, undergoes iterative training involving an offline
feature clustering step and a masked prediction step. 
VatLM extends AV-HuBERT by further integrating the text modality to achieve unified visual-audio-text representation learning~\citep{zhu2023vatlm}.
Another line of work adopts momentum-based teacher to generate learning targets online, such as RAVEn~\citep{haliassos2023jointly}, AV-data2vec~\citep{Lian2023avdata2vec}, and our previously proposed AV2vec ~\citep{Zhang2023av2vec}.
% Audio-visual speech self-supervised learning approaches have achieved impressive performance on lipreading task. 
These methods necessitate a substantial number of training steps for the audio-visual self-supervised pretraining to converge, incurring significant computational resource and time requirements. 
For instance, the base version of the AV-HuBERT model demands 5 iterations, each involving 400,000 training steps~\citep{shiavhubert}.
While AV2vec significantly accelerates the training process without requiring iterations, training on the LRS3 dataset still takes 7 days on 8 Tesla A100 GPUs~\citep{Zhang2023av2vec}. Therefore, transferring a pretrained model from a source language to a target language can yield significant time savings during the training phase.

\begin{figure*}[t]
    \centering
    \includegraphics[width=0.9\textwidth]{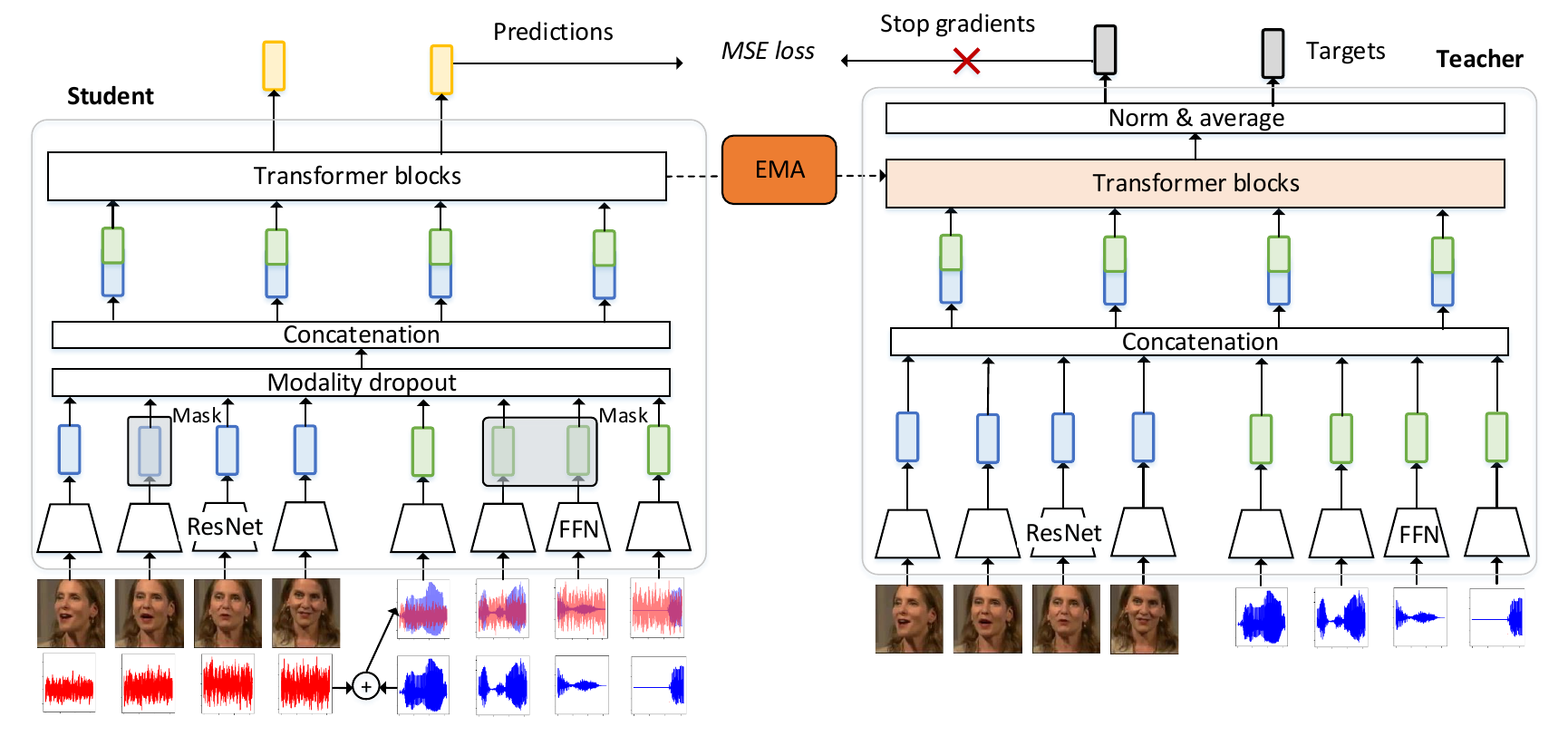}
    \caption{Illustration of AV2vec model. AV2vec contains a student and a teacher module. The student performs a masked latent feature regression task and the targets are generated online by the teacher. The teacher weights are an exponentially moving average (EMA) of the student parameters.}
    \label{fig:figure1}
\end{figure*}

\section{AV2vec}
 AV2vec pretraining aims at learning general audio-visual speech representations by multimodal self-distillation.
AV2vec consists of a student and a teacher module, where the
student is trained with a masked latent feature regression task and the
targets are generated online by the teacher. The model is illustrated
in Figure~\ref{fig:figure1}.

Let the clean audio be $\boldsymbol{A}^c =[\boldsymbol{a}^c_1, \boldsymbol{a}^c_2,\dots, \boldsymbol{a}^c_T]$, where $\boldsymbol{a}^c_{i} \in \mathbb{R}^{D_A}$ represents an acoustic feature frame and $D_A$ is the feature dimension. $T$ represents the number of frames. Let the video input be $\boldsymbol{V} =
[\boldsymbol{v}_1,\boldsymbol{v}_2,\dots,\boldsymbol{v}_T]$, where $\boldsymbol{v}_i \in \mathbb{R}^{H \times W \times C}$ represents a video frame and $H/W/C$ are height/width/channel dimension respectively.
The audio and video streams are synchronized at the same frame rate through audio frame reduction, as detailed in Section~\ref{sec:modelconfig}. 
%The audio and video stream share the same frame rate by using a frame reduction  for audio, as described in detail in Section~\ref{sec:modelconfig}.
The audio input
of the student $\boldsymbol{A} =[\boldsymbol{a}_1, \boldsymbol{a}_2,\dots, \boldsymbol{a}_T]$  is obtained by adding noise to clean audio randomly  with a probability $p_{noise}$. The audio and visual input is processed by modality-specific feature extractors for producing audio intermediate features $\boldsymbol{F}^a = [\boldsymbol{f}^a_1, \boldsymbol{f}^a_2,\dots,\boldsymbol{f}^a_T]$ and visual intermediate features
 $\boldsymbol{F}^v = [\boldsymbol{f}^v_1, \boldsymbol{f}^v_2,\dots,\boldsymbol{f}^v_T]$, respectively, where $\boldsymbol{f}^a_i \in \mathbb{R}^{D_{fa}}$ and $\boldsymbol{f}^v_i \in \mathbb{R}^{D_{fv}}$. $D_{fa}$ and $D_{fv}$ represent the 
 audio and visual feature dimension, respectively. 
 A feed forward network and a 3DCNN-ResNet18~\citep{afouras2018deep} network are adopted for audio and visual feature extraction, which are also referred to as the audio and video frontend, respectively. 
 Then features of each modality are corrupted independently by a span-based mask. 
  Specifically, 
 let $M_a$ be a set of masked indices for audio, $\tilde{\boldsymbol{F}}^a
 = r(\boldsymbol{F}^a, M_a)$ denotes audio feature $\boldsymbol{F}^a$
 is corrupted by replacing $\boldsymbol{f}^a_i$ with an embedding $\boldsymbol{e}^a$ if $i \in M_a$. For corrupting visual features, the same strategy as the audio is used but with different mask indices
 $M_v$ and embedding $\boldsymbol{e}^v$.
 Then, modality dropout 
 is employed to train the student 
 in absence of audio or video stream randomly. Specifically, with a probability
 $p_m$, both modalities are used. The probability of selecting audio is
 $p_a$ when only one modality is used, and features of the other 
 modality are set to all zeros.  Audio-visual representations
 are then obtained as $\boldsymbol{F}^{av} = \text{concat}(\tilde{\boldsymbol{F}}^a, \tilde{\boldsymbol{F}}^v)$,
 where $\text{concat}( \cdot )$ denotes channel-wise concatenation.
 $\boldsymbol{F}^{av}$ is fed into a stack of Transformer blocks for generating
multimodal contextualized features, which are passed through
a linear projection layer for regressing to latent representations generated by the teacher. A mean square error loss is used as 
\begin{equation}
L_{reg} = \sum_{t \in M_a \cup M_v} || \boldsymbol{x}_t - \boldsymbol{y}_t||^2,
\end{equation}
where $\boldsymbol{x}_t$ is the prediction and $\boldsymbol{y}_t$ is the target produced by the teacher. Note that we only penalized the predictions in masked regions.

The teacher module consumes both clean audio $\boldsymbol{A}^c$ and
video $\boldsymbol{V}$ as input. Intermediate feature masking or modality
dropout is not applied for the teacher. Therefore, it generates
hidden representations with full context and multimodal input.

The teacher weights are updated by an exponentially moving 
average (EMA) of the student as
\begin{equation}
\boldsymbol{\theta}_i = \lambda \cdot \boldsymbol{\theta}_{i-1} + (1-\lambda) \cdot \boldsymbol{\phi}_{i},
\end{equation}
where $\boldsymbol{\theta}$ and $\boldsymbol{\phi}$ denote parameters of the teacher and the student, respectively, $i$ denotes the update step and $\lambda$ is a hyper-parameter that controls the update step.
$\lambda$ is linearly increased from $\lambda_b$ to $\lambda_e$
at the initial ${n}_{warmup}$ updates and it is kept constant in the rest
training steps.
This momentum update is 
applied only for the Transformer blocks and the teacher's feature extractors share
the parameters with the student. No gradients flow to the teacher 
and only the student is trained with the back-propagation algorithm.
For obtaining targets of the student, instance normalization is applied for 
the last $k$ Transformer blocks' hidden outputs and then they are
averaged. 
Normalizing the targets helps prevent the model from collapsing into a constant representation for all time-step and averaging multi-layer
representations boosts the performance of learning.

\begin{figure*}[t]
    \centering
    \includegraphics[width=0.75\textwidth]{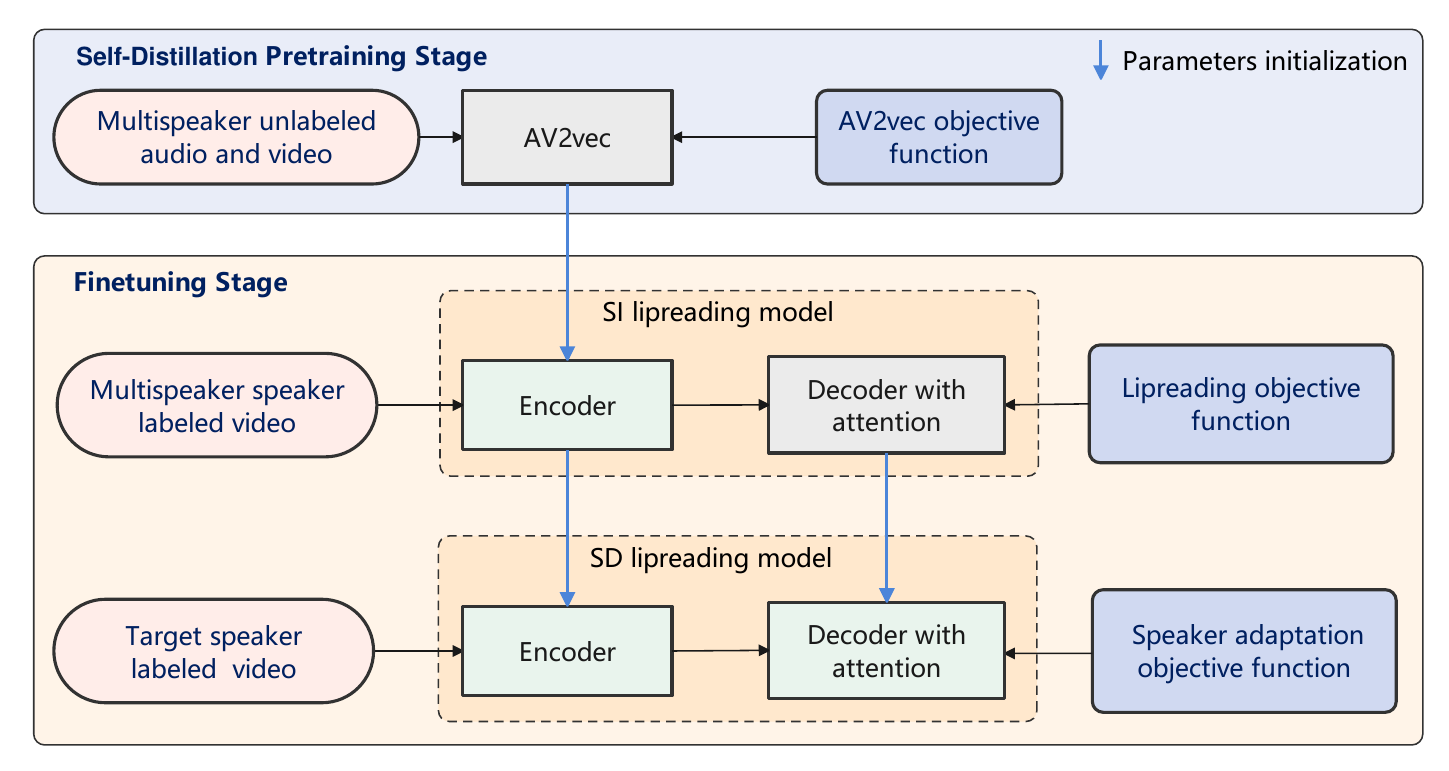}
    \caption{Illustration of the training scheme of our proposed method for the target speaker lipreading. ``SI'' and ``SD" represent speaker-independent and speaker-dependent respectively.}
    \label{fig:figure2}
\end{figure*}

\section{Proposed method}

Our method consists of 3 phases during training as illustrated in Figure~\ref{fig:figure2}.
At the pretraining stage, the AV2vec model is pretrained by self-distillation on parallel audio visual data without using any text labels. Then, the pretrained AV2vec model is employed as the encoder for extracting visual representations and a randomly initialized decoder is  adopted for decoding texts. The encoder-decoder model is jointly optimized for the lipreading task during finetuning. A speaker-independent (SI) model is first finetuned with videos from a diverse range of speakers. After that, the speaker-dependent (SD) model, whose parameters are initialized by the SI model, is further finetuned with only the video data from the target speaker.

In the study, we examine two different approaches for the visual input of the model: the conventional lip region-of-interest (ROI) and the entire face, comparing their efficacy through experimental analysis. 
To procure videos of the lip ROI, we initially detect and align 68 facial keypoints to a reference face using an affine transformation. A 96 × 96-pixel region, centered on the mouth, is then extracted from the resized 224 × 224-pixel image of the entire face. This cropped area undergoes data augmentation, including random cropping and horizontal flipping, and it is subsequently converted to grayscale and normalized to serve as the final visual input for the model.
For the generation of full-face videos, images from the training dataset are resized to 96 × 96 pixels. These images are similarly subjected to data augmentation, converted to grayscale, and normalized. 
Regarding the audio input of the AV2vec pretraining, log filterbank features are utilized. 

\subsection{Speaker-independent model}

\textbf{Joint CTC/attention architecure}: To develop a speaker-independent lipreading model, a Transformer-based decoder with 
cross-attention is integrated atop the AV2vec encoder. 
% The audio branch of the AV2vec is
% discarded. 
The audio branch of the AV2vec is omitted. This encoder-decoder model is then fine-tuned on multi-speaker video data accompanied by their corresponding text transcriptions.
The primary training objective is to minimize 
the overall negative log probability of the label sequence, which is
factorized using the chain rule of probability as:
\begin{equation}
L_{ce} = - \frac{1}{N}\sum_{n=1}^{N} \log(\text{P}(t_{n} |\boldsymbol{t}_{<n}, \boldsymbol{V})),
\end{equation}
where $N$ represent the text sequence length.
$\boldsymbol{t} = [t_{1}, \dots, t_{N}]$ 
represents the text label sequence from a vocabulary $\mathcal{U}$, i.e. $t_{i} \in \mathcal{U}$. $P(:|\boldsymbol{t}_{<n}, 
\boldsymbol{V})$ represents the conditional probability distribution predicted by the model given $n$ historical elements in the label sequence and video input. 
This loss is also usually referred to as a cross entropy loss between
the predicted  probability distribution across the output vocabulary and the ground truth labels.

In addition to the cross entropy loss, 
the connectionist temporal classification (CTC) loss
is also integrated to train the lipreading model, resulting in
a hybrid CTC/attention architecture~\citep{watanabe2017hybrid}. Specifically, 
a linear projection layer is employed, utilizing the encoder representations
as input for predicting the targets from an
augmented vocabulary that includes a special blank symbol $\langle b \rangle$, denoted as $\mathcal{U}' = \mathcal{U} \cup \langle b \rangle$. 
%$T$ represents the length of encoder output.
Consider $\hat{\boldsymbol{t}}_{1:T}$ as one of the potential label sequences in CTC computation and  $\boldsymbol{t}_{1:N}$ as the ground truth text sequence.
A mapping $\mathcal{B}(\hat{\boldsymbol{t}}_{1:T}) = \boldsymbol{t}_{1:N}$ is defined, where
$\hat{\boldsymbol{t}}_{1:T}$ is considered equivalent to $\boldsymbol{t}_{1:N}$ after first removing consecutive identical symbols, and then eliminating any blank symbols.
The CTC loss is defined as follows:
\begin{equation}
L_{ctc} = -\log \left( \sum_{\hat{\boldsymbol{t}}_{1:T} \in \mathcal{B}^{-1}(\boldsymbol{t}_{1:N})} 
\prod_{i=1}^T
\text{P}_{ctc}(\hat{t}_i|\boldsymbol{V}) \right),
\end{equation}
 where $\mathcal{B}^{-1}$ represents the inverse function of $\mathcal{B}$.
The joint architecture shares the same encoder but uses
separate mechanisms in the decoder, which can be considered as multitask learning.
During training, the objective function is performed by a combination of the CTC and attention objectives, which is computed as follows:
\begin{equation}
L_{si}= L_{ce} + \mu L_{ctc},
\end{equation}
where $\mu$ is the weight of the CTC loss during joint training.

\textbf{Transfer learning:}
A cross-lingual transfer learning approach is proposed for the lipreading task. Specifically, the AV2vec model, pretrained on a source language, is finetuned
with the hybrid CTC/attention loss for lipreading in a target language. The AV2vec model is expected to learn cross-lingual shared visual representations, enabling knowledge transfer to the target language. By utilizing
an already pretrained model in a source language,
this transfer learning strategy not only mitigates the high computational costs associated with audio-visual self-supervised pretraining, but it also facilitates performance improvements when target language data is limited. During the fine-tuning process, the encoder model parameters are initially frozen for the first $k$ steps and subsequently jointly optimized with the decoder. 
However, for the transfer learning of AV2vec, the encoder parameters are not frozen,
allowing for continuous optimization alongside the decoder.

%In contrast, for AV2vec transfer learning, the encoder parameters are kept trainable from the start, enabling better adaptation to the visual-phonetic patterns of the target language.
%However, for the transfer learning of AV2vec, the encoder parameters are not frozen, allowing more sufficient adaptation for modeling videos of  the target language .

\subsection{Speaker-dependent model}

The speaker-independent model is trained for improving average lipreading accuracy
across various speakers. 
At this stage, the lipreading model is further optimized for
a specific target speaker. 
Only the video data of the target speaker is adopted for speaker adaptation and the model parameters are updated with a small learning rate.
However, the lipreading model may still suffer from the overfitting effect because
of the typically small amount of training data available for the target speaker.
To address this issue, drawing inspiration from the speaker adaptation techniques used in speech recognition, an additional KL-divergence (KLD) loss is introduced. 
This loss measures the difference between the predicted probability distributions of the speaker-independent model and the speaker-dependent model as:
\begin{equation}
%\begin{split}
L_{kld} = \frac{1}{N} \sum_{n=1}^{N} \sum_{i=1}^{|\mathcal{U}|} \text{P} (i | \boldsymbol{t}_{<n}, \boldsymbol{V}; \boldsymbol{\theta}_{SI}) %*  \\
\log \left(
\frac{\text{P}(i | \boldsymbol{t}_{<n}, \boldsymbol{V}; \boldsymbol{\theta}_{SI})}{\text{P}(i | \boldsymbol{t}_{<n}, \boldsymbol{V};\boldsymbol{\theta}_{SD})}
\right),
%\end{split}
\end{equation}
where $\boldsymbol{\theta}_{SI}$ and  $\boldsymbol{\theta}_{SD}$ represent the parameters of the speaker-independent and 
speaker-dependent model, respectively.

The total loss for speaker adaptation is a weighted sum 
of the original cross-entropy loss and the KLD loss term, which is controlled by
a regularization weight $\rho$ as:
\begin{equation}
L_{sa} = (1-\rho) L_{ce} + \rho L_{kld} + \mu L_{ctc}. 
\end{equation}
By omitting the untrainable speaker-independent term related to the loss function,
the speaker adaptation loss can be simplified as:
\begin{equation}
%\begin{split}
L_{sa} = - \frac{1}{N} \sum_{n=1}^{N}
                        \sum_{i=1}^{|\mathcal{U}|}
                        ((1-\rho)\delta(t_n-i) + \rho \text{P} (i | \boldsymbol{t}_{<n}; \boldsymbol{\theta}_{SI})) %* \\
 \log \text{P} (i | \boldsymbol{t}_{<n}; \boldsymbol{\theta}_{SD}) +\mu L_{ctc},
 %\end{split}
\end{equation}
where $\delta(i)=1$ if $i=0$, otherwise $\delta(i)=0$.
The first term of this loss can be interpreted as the cross-entropy loss using
the interpolation of the one-hot encoded label and the predicted probability distribution of the speaker-independent model.

\subsection{Model ensemble in testing}

Ensemble learning~\citep{polikar2012ensemble} aims to improve overall model performance by aggregating predictions from multiple models. We adopt this approach by averaging the probability distributions from lipreading models using either full facial images or lip region-of-interest (ROI) inputs. 
A joint CTC/attention decoding approach~\citep{watanabe2017hybrid}
is employed, which alleviates the drawback of the attention-only approach that has non-monotonic alignment and end-of-sentence detection issues.
The decoding process is described by the following equation:
\begin{equation}
\begin{split}
\boldsymbol{t}^* = \text{argmax}_{\boldsymbol{t}} \Bigg\{ \frac{1}{M} \sum_{i=1}^{M}( (1-\alpha) \text{P}(\boldsymbol{t}|\boldsymbol{V}_{lip} \ \text{or} 
 \ \boldsymbol{V}_{face}; \boldsymbol{\theta}_i)) \\ +
 \alpha \text{P}_{ctc}(\boldsymbol{t}|\boldsymbol{V}_{lip} \ \text{or} 
 \ \boldsymbol{V}_{face}; \boldsymbol{\theta}_i) \Bigg\},
\end{split}
\end{equation}
where $M$ represents the number of base models and  $\alpha$ is the weight of CTC. $\boldsymbol{V}_{lip}$
and $\boldsymbol{V}_{face}$ represent video of the lip ROI and the entire face, respectively. 
In addition to simple averaging of model predictions, 
we can optimize individual model weighting coefficients to enhance ensemble performance. 
However, we opted for equal weighting based on its good performance in our preliminary experiments.

%Besides equally averaging
%each model's predictions, we can also optimize the weighting coefficients for individual models to enhance the overall ensemble performance.
%We used equal weighting because it produced the good results in our preliminary experiments.

While ensemble learning is a conventional technique for boosting performance, our key contribution lies in applying it to lipreading models leveraging different input types: full facial regions or lip ROIs. Our previous work~\citep{Zhang2022face} showed that using full facial images outperforms lip ROIs by providing additional visual cues beneficial for lipreading. 
However, employing the entire facial region as input rather than solely focusing on the lip area can introduce extraneous visual distractions, such as hand movements near the face or background elements in the facial video. 
These irrelevant factors may degrade the performance of lipreading models in certain scenarios.
Conversely, the lip ROI input is more robust to variations in the video background. By ensembling models with different types of inputs, the overall lipreading performance can be enhanced due to their complementary nature.

\section{Experiments and results}

\begin{table}[t]
    \centering
    \begin{tabular}{cccccc}
    \hline
    \hline
      \textbf{Dataset}  &  \textbf{Lang} &  \textbf{Split} & \#\textbf{Spk} & \#\textbf{Utt}  & \#\textbf{Hour} \\
      \hline
        \multirow{3}{*}{LRS3} & \multirow{3}{*}{EN} & Pre-train &  5090 & 119k &  407\\ 
         &  & Train-val & 4004 & 32k &  30 \\
         &  & Test & 451 & 1452 & 1 \\
         \hline
        \multirow{2}{*}{CN-CVS} & \multirow{2}{*}{CH} & News  & 28 & 13k & 35 \\
         &  & Speech & 2529 & 193k &  273 \\
         \hline
         \multirow{3}{*}{ChatCLR} & \multirow{3}{*}{CH} & Train &  229 & 173k & 111 \\
         & & Dev & 12 & 5722 & 4.5 \\
         & & Eval & 12 & 3081 & 2.4 \\
         \hline
         \hline
    \end{tabular}
    \caption{Datasets utilized in our experiments. ``EN'' and ``CH" represent English and Chinese respectively. ``ChatCLR" represents
    the official dataset for the task 2 of 2024 ChatCLR Challenge, of which training split is obtained by combining the training and development set of the MISP2021 AVSR dataset's far-field videos~\citep{misp2021}. 
    }
    \label{tab:tabel1}
\end{table}

\subsection{Data}

In our experiments, we focused and evaluated our proposed methods on the task 2 (i.e., target speaker sentence-level lipreading) of the 2024 Chat-scenario Chinese Lipreading (ChatCLR) Challenge\footnote{https://mispchallenge.github.io/ICME2024/}, which is a sentence-level lipreading task involving the target speaker. 
The dataset was recorded in realistic home TV scenarios, capturing colloquial conversation in Chinese.
The videos were recorded by a far-field camera in a realistic home TV setting, resulting in substantial variations in speakers' head poses, illumination conditions, and emotional expressions.
Also, the speakers' manner of speaking is highly casual and colloquial, deviating from formal speech patterns.

For training models, three audio-visual speech datasets were utilized in our experiments, including LRS3~\citep{Afouras18d}, CN-CVS~\citep{chen2023cn}, and ChatCLR\footnote{https://mispchallenge.github.io/ICME2024/data.html},
with their detailed specifications summarized in Table~\ref{tab:tabel1}.
The LRS3 dataset comprises approximately 438 hours of video extracted from TED and TEDx talks in English, sourced from YouTube.
The pre-train and train-val splits of this dataset were adopted for the AV2vec pretraining in our experiments.
The CN-CVS dataset contains about a total of 308 hours of Mandarin videos, which contains a News part collected from TV news with frontal cameras and a Speech part from speech shows captured with flexible cameras.
Both parts of the CN-CVS dataset were combined  for  pretraining AV2vec 
and  finetuning speaker-independent models.
The ChatCLR dataset is the officially constructed dataset by the Challenge organizer. The training part of
this dataset contains 111 hours Mandarin videos, which was compiled by merging the training and development sets of the MISP2021 AVSR dataset’s far-field videos~\citep{misp2021}.
%For audios, 
The MISP2021 dataset includes audio recordings captured simultaneously from near-field, middle-field, and far-field microphone placements. 
For AV2vec pretraining, only the near-field recordings were used.
%The MISP2021 dataset features audio recordings captured simultaneously at three spatial positions: near-field, middle-field, and far-field microphone placements. 
%For AV2vec pretraining, we only utilized the near-field recordings.
%MISP2021 provides three types of audio recorded simultaneously by near-field, middle-field, and far-field micrphones respectively.
%We utilized only the near-field audio for AV2vec pretraining.
%For AV2vec pretraining, only speech recorded by near-field microphones in MISP 2021 dataset was used. 
The MUSAN dataset\footnote{https://www.openslr.org/17/} was employed to add noise to the audio inputs.
Both the development and evaluation sets of the ChatCLR dataset feature the same 12 target speakers, with an equal gender split of 6 males and 6 females. 
Utterances in the development set were randomly split in an 8:2 ratio for training and validation in speaker adaptation. Our methods were also tested on the official ChatCLR evaluation set, and the results were reported accordingly.

\subsection{Model configuration}
\label{sec:modelconfig}
For the video input of the model,  $88 \times 88 $-pixel images were randomly cropped
from the pre-processed $96 \times 96$-pixel images of the lip ROI or the entire face following our previous study~\citep{Zhang2023av2vec}.
These images were then randomly flipped horizontally with a probability of 0.5 during training. At test time, 88 × 88-pixel images were center cropped and did not undergo horizontal flipping. For the audio input of the AV2vec model, 26-dimensional log filter bank features were extracted every 10 milliseconds, and every four consecutive frames were stacked in the channel dimension to synchronize the frame rate with that of the video (i.e., 25 fps).

The details of our model architecture are presented in~\ref{app1}, which followed our previous study~\citep{Zhang2023av2vec}. 
For the audio and visual frontends, a single fully connected layer and a 3D CNN-ResNet were utilized, respectively.  The encoder consisted of 12 Transformer blocks~\citep{vaswani2017attention} equipped with pre-layer normalization~\citep{xiong2020layer}. 
After pretraining, a decoder was further added on the top of the encoder.
This decoder comprised 6 Transformer blocks equipped with the additional cross attention component to 
process the representations from the encoder. 

The AV2vec pretraining configuration follows the same hyperparameter settings
established in our prior work~\citep{Zhang2023av2vec}.
%The configuration for AV2vec pretraining is inherited from our previous work~\citep{Zhang2023av2vec}. 
Specifically, the probability of noise $p_{noise}$ was set to 0.25. 
80\% of audio features
and 30\% of video features were masked for the student. $p_m$ and
$p_a$ were set to 0.5 and 0.5, respectively.
$\lambda_b$, $\lambda_e$, and  ${n}_{warmup}$ were set to 0.999, 0.9999, and 30k, respectively.
The last 8 layers were averaged representations of the for the teacher. 
The Adam optimizer was used with a peak learning rate of $5 \times 10^{-4}$. The total number of the updates during pretraining was 400k. The learning rate was first warmed up linearly for the first 3\% of the updates and then kept constant 
for 90\% of the updates followed by exponentially 
decay over the remaining 7\%.
During the finetuning of the speaker-independent model , a vocabulary comprising a total of 6000  Chinese characters and English subwords
was utilized. 
The Adam optimizer was employed again with a peak learning rate of $1 \times 10^{-3}$. The learning rate was first warmed up linearly for the initial 40k of updates and then exponentially 
decayed over the remaining 80k updates. The encoder was frozen for the initial 60k updates except for transfer learning.
During the finetuning of speaker-dependent model, the peak learning rate was set to $1 \times 10^{-4}$. The learning rate was first
warmed up linearly for the first 200 updates and then exponentially 
decayed over the remaining 800 updates. 
The hyper-parameter $\rho$ was set to 0.1 during speaker adaptation.
During training and decoding, the weights of the CTC, represented by
$\mu$ and $\alpha$ respectively, were both set to 0.1.
During decoding, beam search was adopted with a beam size of 50.
Our validation set was used to optimize the hyperparameters during finetuning.

Four types of models were trained in our experiments and they are described as follows:

\begin{itemize}
    \item \textbf{AV2vec-lip}: 
    For both the AV2vec pretraining and speaker-independent finetuning stages, this method utilized datasets in Chinese, namely the ChatCLR and
    the CN-CVS corpus.
    This model adopted the traditional lip ROI as visual input.
\item \textbf{AV2vec-face}: This method had the same configuration as AV2vec-lip except using the entire face as visual input.

\item \textbf{AV2vec-tf-lip}: The model underwent pretraining on an English source language dataset, namely the LRS3 corpus. Subsequently, during the finetuning stage, transfer learning was applied to adapt the model for lipreading in the target Chinese language. For the speaker-independent model finetuning experiments, the ChatCLR and CN-CVS datasets were utilized.

\item \textbf{AV2vec-tf-face}: This method had the same configuration as AV2vec-tf-lip except that it utilized the entire face as visual input.
\end{itemize}

Character error rate (CER) of recognition results was employed as
the evaluation metric in our experiments. It is computed by
the edit distance between the model predicted texts and the 
ground truth labels as:
\begin{equation}
    \text{CER} = \frac{S+ I+ D}{N} \cdot 100\%,
\end{equation}
where $S$, $I$, and $D$ represent the number of substitution, insertion, and deletion errors respectively. $N$ represents the total number of label characters.

In our experiments, our method was primarily evaluated on our internal validation set, as the labels for the official evaluation set were kept confidential from competitors. The performance of our approach on the official evaluation set of the ChatCLR Challenge was subsequently reported and compared against other teams, which is presented in Section~\ref{sec:subsubChall} of our experimental results.

Our method was implemented with the Fairseq framework\footnote{https://github.com/facebookresearch/fairseq}. 8 Tesla A100 GPUs were adopted in our experiments. 
The AV2vec pretraining cost around 7 days to converge. 
The finetuning phase took approximately 17 hours for the non-transfer models (AV2vec-lip and AV2vec-face) 
and 22 hours for the transfer models (AV2vec-tf-lip and AV2vec-tf-face). 
 An additional 30 minutes of speaker adaptation was required per speaker to obtain the speaker-dependent models.

%or speaker-dependent models, the fine-tuning process required 30 minutes per speaker.
%For the non-transfer models (i.e., AV2vec-lip and AV2vec-face) and tranfer models (i.e., AV2vec-tf-lip and AV2vec-tf-face), 
%finetuning phase cost about 22 hours and 17 hours, respectively.
%The speaker-dependent model finetuning cost
%0.5 hours for each speaker. 

\subsection{Experimental results}

\begin{table}
    \centering
    \scalebox{0.9}{
    \begin{tabular}{cccc}
    \hline
    \hline
        Method & PT data & FT data & CER(\%)$\downarrow$ \\
    \hline
        Afouras et al.~\citep{afouras2018deep} & -- & LRS2,3+MVLRS &  58.9 \\
        Ma et al.~\citep{ma2021end} & -- & LRS3+LRW & 43.3 \\
        \multirow{2}{*}{AV-HuBERT~\citep{shiavhubert}} & \multirow{2}{*}{LRS3} & LRS3-30h & 47.1 \\
                                    &  & LRS3 & 40.3 \\
         \multirow{2}{*}{REVAn~\citep{haliassos2023jointly}} & \multirow{2}{*}{LRS3} & LRS3-30h & 47.0 \\
                                    &  & LRS3 & 39.1 \\
      \multirow{2}{*}{AV-data2vec~\citep{Lian2023avdata2vec}} & \multirow{2}{*}{LRS3} & LRS3-30h & 45.2 \\
                                    &  & LRS3 & 39.0 \\
        \hline
         \multirow{2}{*}{AV2vec-lip} &  \multirow{4}{*}{LRS3} & LRS3-30h &  45.1 \\ 
          &  & LRS3 &  39.9  \\
         \multirow{2}{*}{AV2vec-face} &  & LRS3-30h & 39.4 \\
          &   & LRS3 & 34.3  \\
    \hline 
    \hline
    \end{tabular}
    }
    \caption{Lipreading results on LRS3 test set with AV2vec pretraining. 
     ``LRS3-30h'' represents a 30-hour subset of training data. ``PT'' and ``FT'' represent pretraining and finetuning respectively.}
    \label{tab:table3}
\end{table}

\begin{table}[t]
    \centering
    \scalebox{1.0}{
    \begin{tabular}{cccc}
    \hline
    \hline
        Method & PT data & FT data & CER(\%)$\downarrow$  \\
    \hline
         Supervised-lip &  -- & \multirow{6}{*}{Chat+CN}  & 96.0 $\pm$ 1.4 \\
         %\hline
         AV2vec-tf-lip & LRS3 &  &  82.9  $\pm$  1.6 \\ 
         AV2vec-lip & Chat+CN &   & 75.0  $\pm$ 1.6 \\
         Supervised-face & -- &   & 105.2 $\pm$ 1.1 \\
         %\hline
          AV2vec-tf-face & LRS3 &   & 80.7 $\pm$ 1.7 \\
         AV2vec-face & Chat +CN  &   &  70.4 $\pm$ 1.8 \\
    \hline 
    \hline
    \end{tabular}
    }
    \caption{Lipreading results on ChatCLR validation set for speaker-independent (SI) models.
    % ``TM'' denotes the training time cost on 8 Tesla A100 GPUs. 
    $\pm$ indicates 95\% confidence intervals computed by bootstrap estimation~\citep{bisani2004bootstrap}.
    ``Supervised" represents the model was trained with only supervised learning. 
    %without using any self-supervised learning based pretraining. 
    ``PT'' and ``FT'' represent pretraining and finetuning respectively. ``Chat+CN'' represents the ChatCLR and the CN-CVS datasets.}
    \label{tab:table4}
\end{table}

\subsubsection{Speaker-independent model}

Table~\ref{tab:table3} presents the lipreading performance of the AV2vec models on the LRS3 dataset~\citep{Zhang2023av2vec} and compares it to other recently proposed baseline methods.
For comparability, results from the ``Base'' versions of AV-HuBERT, RAVEn, and AV-data2vec were reported. 
Table~\ref{tab:table3}  demonstrates that the AV2vec method achieved comparable or superior lipreading performance when using the lip region-of-interest (ROI) as the visual input. When employing full face images as the visual input, the character error rate (CER) of lipreading was significantly reduced compared to using the conventional lip ROI.

The lipreading results on the ChatCLR validation set are presented in 
Table~\ref{tab:table4}. As we can see from Table~\ref{tab:table4},
the models that did not employ self-supervised pretraining, specifically Supervised-lip and Supervised-face, recorded
high CER in the lipreading task on our validation set. Notably, Supervised-face exhibited even worse
performance than Supervised-lip, suggesting
that the 
additional information from the face
might detract from lipreading accuracy in the 
absence of self-supervised learning. 
This phenomenon was also
observed in our previous work~\citep{Zhang2022face}.
The AV2vec-tf-lip and AV2vec-tf-face methods were derived from cross-lingual transfer learning based on the English-pretrained AV2vec model listed in Table~\ref{tab:table3}, and subsequently finetuned for Chinese lipreading.
They achieved significantly better results than randomly initialized models, namely Supervised-lip and Supervised-face. These results
demonstrate the effectiveness of our proposed cross-lingual transfer learning for Chinese lipreading.
However, when compared to AV2vec models that underwent pretraining directly on Chinese datasets, the transfer learning-based models exhibited relatively lower performance. This is an expected outcome, as models can learn more effectively when the pretraining and finetuning data are from the same language domain and distribution.
Furthermore, as evident from Table~\ref{tab:table4}, models that leverage AV2vec pretraining tend to benefit from adopting full face inputs, regardless of whether cross-lingual transfer learning is applied or not.

\begin{figure}[t]
    \centering
    \subfigure[Lip ROI-based models]{
    \begin{minipage}[b]{0.46\textwidth}
    \includegraphics[width=\textwidth]{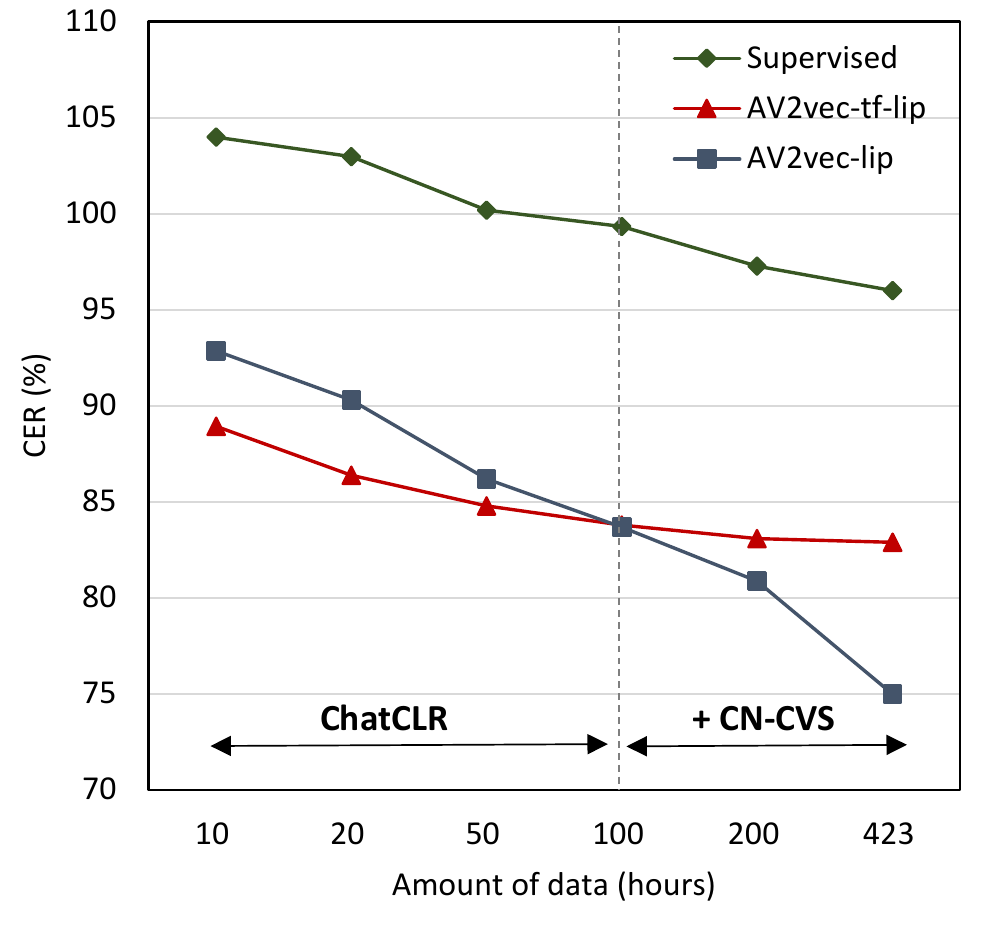}
    \end{minipage}}
     \subfigure[Face-based models]{
    \begin{minipage}[b]{0.46\textwidth}
    \includegraphics[width=\textwidth]{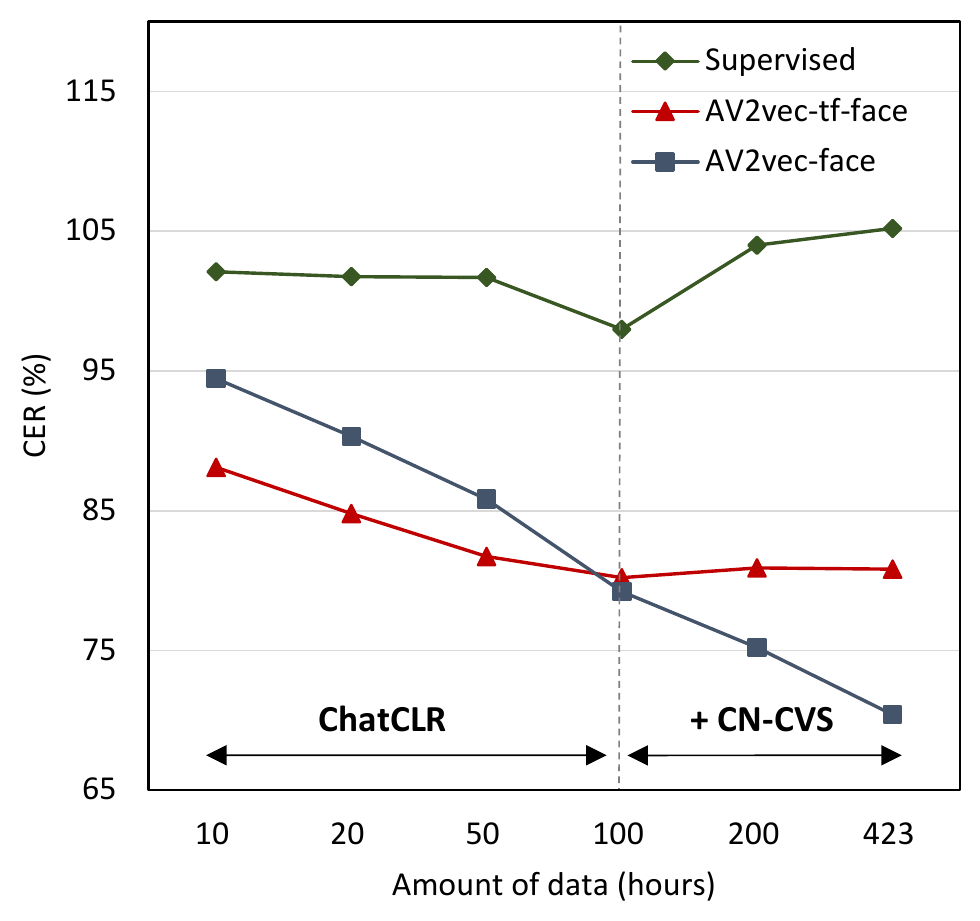}
    \end{minipage}}
    \caption{Lipreading results on ChatCLR validation set when varying
 the amount of training data in Chinese.}
    \label{fig:figure45}
\end{figure}

This section also investigates the scenario where the target language (i.e., Chinese) video data is limited. To simulate this,
datasets containing 200, 100, 50, 20, and 10 hours of Chinese video data were constructed by progressively removing videos first from the CN-CVS corpus and then from the ChatCLR corpus. The lipreading results are summarized in Figure~\ref{fig:figure45}.
As can be observed from Figure~\ref{fig:figure45}, when utilizing the full 423-hour training set, the AV2vec-tf-lip model achieved a higher character error rate (CER) than the AV2vec-lip model. However, the performance gap between AV2vec-tf-lip and AV2vec-lip diminished as the amount of available Chinese video data decreased. Notably, when only 10 hours of Chinese data were available, AV2vec-tf-lip obtained a CER of 88.9\%, outperforming AV2vec-lip, which had a CER of 92.9\%.
A similar trend emerged for the AV2vec-tf-face and AV2vec-face models. Specifically, with only 10 hours of data, the cross-lingual transfer learning model AV2vec-tf-face yielded a CER of 94.5\%, while the non-transfer model AV2vec-face achieved a lower CER of 88.1\%.

For the results of supervised learning models depicted in Figure~\ref{fig:figure45}, an intriguing pattern emerges regarding the character error rate (CER) as the amount of training data increases. The lip region-of-interest (ROI) based method exhibits a consistent decline in CER with more training data. In contrast, the face-based method's CER decreases until the 100-hour mark, after which it rises when additional data is incorporated.
It is noteworthy that the CN-CVS dataset, collected from a domain distinct from ChatCLR, is only present in the 200-hour and full 423-hour training subsets. These results indicate that employing lip ROI can effectively mitigate the domain shift across videos captured under varying conditions. Consequently, incorporating videos from the CN-CVS corpus, which originates from domains different from the ChatCLR validation set, into the training process yields a more robust lipreading model.
Conversely, when leveraging self-supervised AV2vec pretraining, both the lip ROI and face-based methods exhibit performance gains from the additional CN-CVS data, with the exception of AV2vec-tf-face. This finding suggests a higher degree of robustness in self-supervised learning techniques for extracting visual representations across different domains, compared to their supervised counterparts.

\subsubsection{Speaker-dependent model}

\begin{figure*}
\centering
    \subfigure[AV2vec-tf-lip]{
    \begin{minipage}[b]{0.47\textwidth}
    \includegraphics[width=\textwidth]{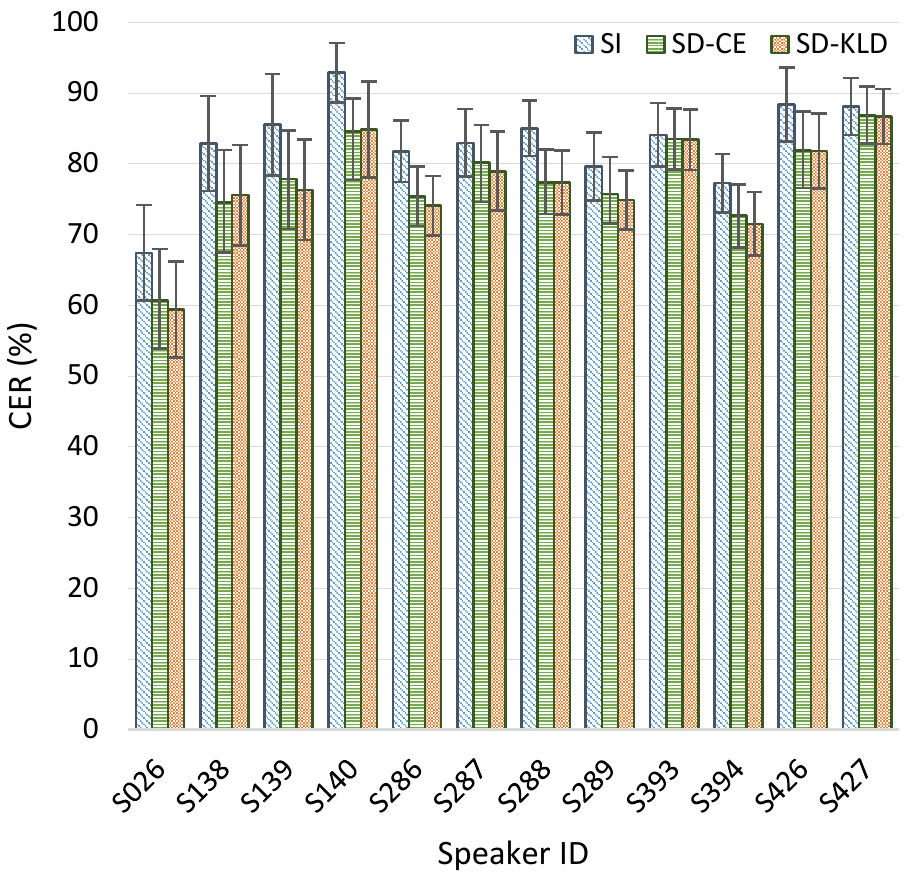}
    \end{minipage}}
    \subfigure[AV2vec-tf-face]{
    \begin{minipage}[b]{0.47\textwidth}
    \includegraphics[width=\textwidth]{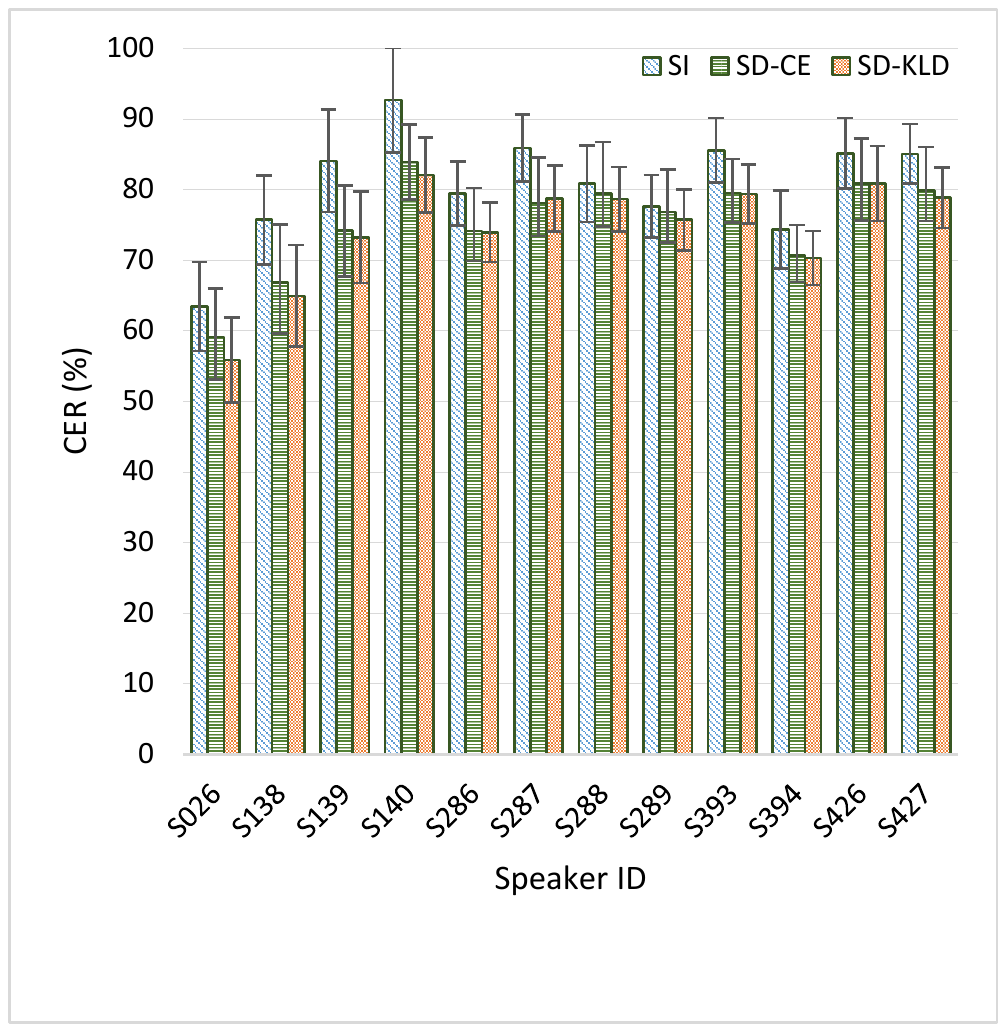}
    \end{minipage}}

\subfigure[AV2vec-lip]{
     \begin{minipage}[b]{0.47\textwidth}
    \includegraphics[width=\textwidth]{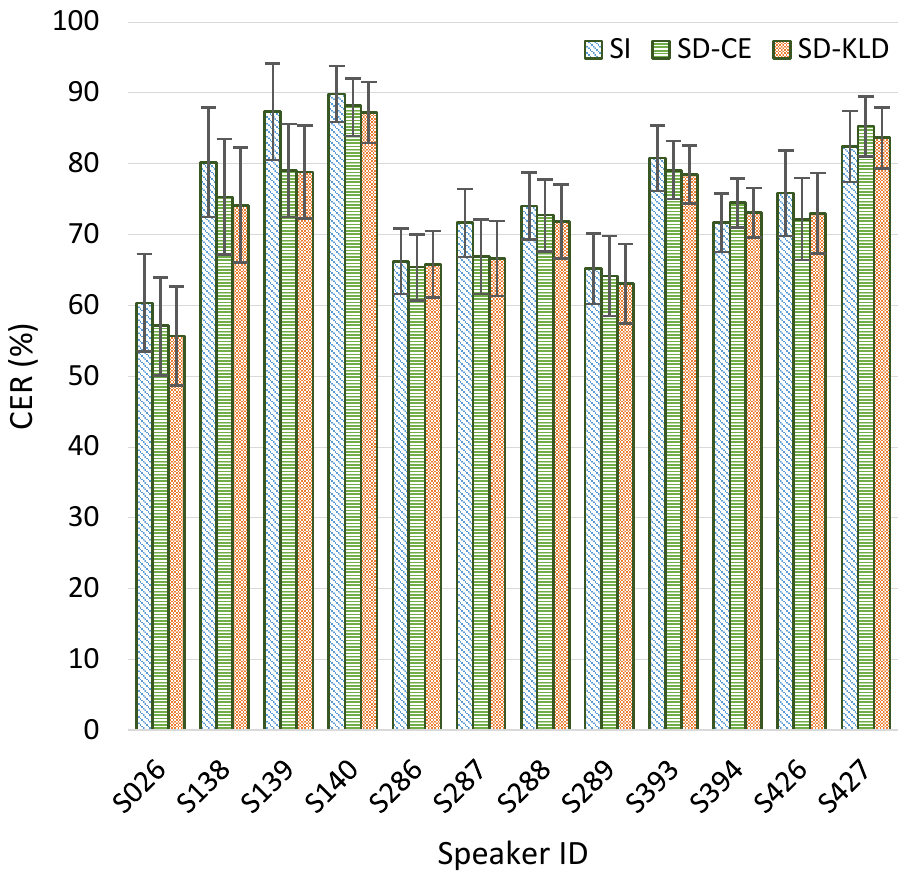}
    \end{minipage}}   
\subfigure[AV2vec-face]{
     \begin{minipage}[b]{0.47\textwidth}
    \includegraphics[width=\textwidth]{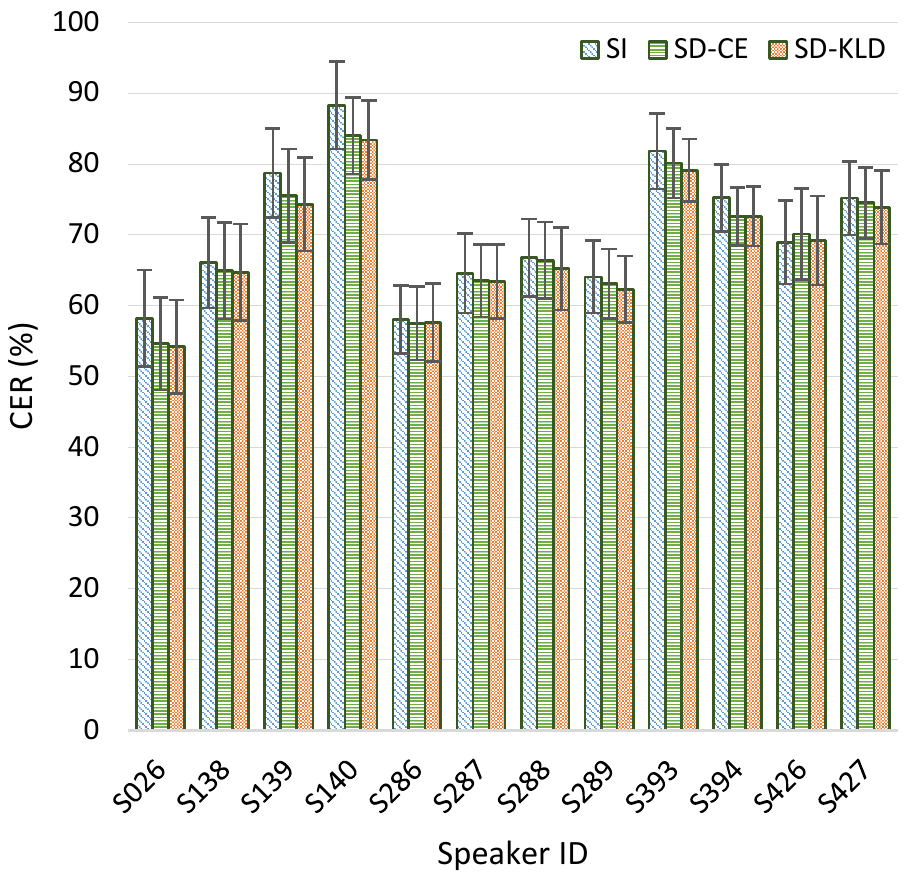}
    \end{minipage}}
\caption{Lipreading results on ChatCLR validation set before and after speaker adaptation. Error bars indicate 95\% confidence intervals. ``SI'' and ``SD'' represent speaker-independent and speaker-dependent, respectively. ``CE'' and ``KLD'' represent using only cross-entropy loss and using additional KLD regularization loss during speaker adaptation, respectively.}
    \label{fig:fig6}
\end{figure*}

After obtaining speaker-independent (SI) lipreading models, we further finetuned them on each of the 12 target speakers respectively for improving the
lipreading accuracy on the each target speaker.
As described in Section 4.2, KLD loss term was used for regularizing the model 
during speaker adaptation. Models without utilizing KLD loss were also constructed
and compared to those with KLD regularization.
Our experimental results are presented in Figure~\ref{fig:fig6}.

Figure~\ref{fig:fig6} reveals substantial speaker-dependent variations in lipreading accuracy. While speaker-independent models initially showed high error rates, speaker adaptation techniques effectively reduced CER. Among the speaker-dependent approaches, AV2vec-face achieved the best performance with an average 68.4\% CER on the ChatCLR validation set, followed by AV2vec-lip (72.6\%), AV2vec-tf-face (74.6\%), and AV2vec-tf-lip (77.2\%), as presented in Table~\ref{tab:tab5}.
Models implementing KLD regularization (SD-KLD) outperformed their non-regularized counterparts (SD-CE) across most target speakers. Statistical analysis with 95\% confidence intervals showed significant improvements after speaker adaptation for AV2vec-tf-lip and AV2vec-tf-face methods, though non-transfer models showed no significant changes. While individual speaker analysis revealed mostly non-significant CER improvements as shown in Figure~\ref{fig:fig6}, these results should be considered within the context of our limited validation set (a total of 1,140 utterances, 14,256 characters for 12 speakers), which led to substantial random effects and wide confidence intervals for individual speaker.

\subsubsection{Model ensemble}

\begin{figure*}
    \centering
    \includegraphics[width=1\linewidth]{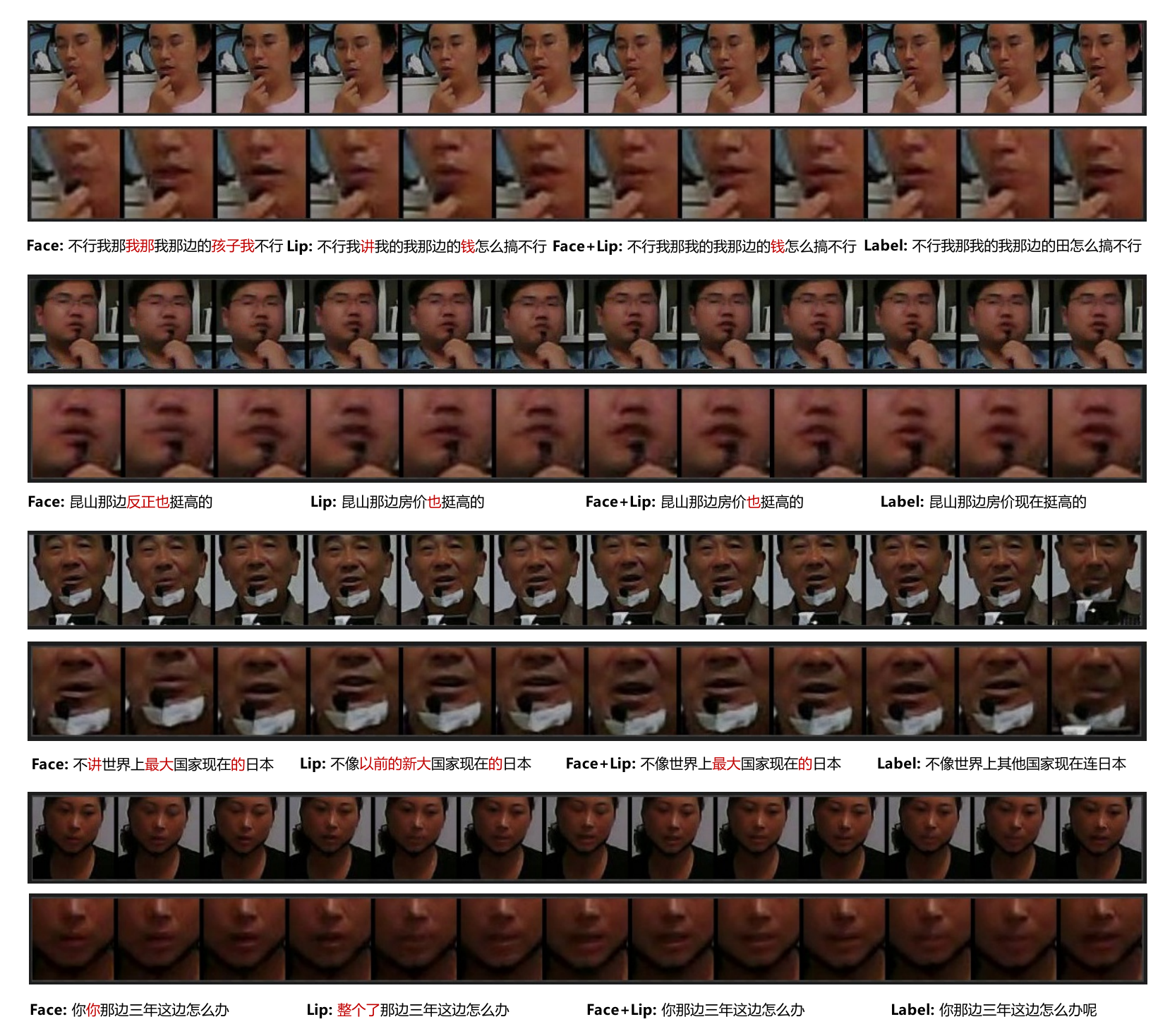}
    \caption{Examples of full face and corresponding lip ROI video frames, along with the lipreading recognition results.  ``Face+Lip'' represents the ensembled model that integrates
    both face and lip ROI inputs. Red characters highlight insertion or substitution errors.}
    \label{fig:figure7}
\end{figure*}

While facial input generally enhances lipreading performance over using only the lip ROI, we identified a scenario where facial input can be detrimental. 
When speakers hold a microphone near their mouth, as shown in the upper part of Figure~\ref{fig:figure7}, the hand movements can occlude the lips and introduce additional motion artifacts that confound lipreading models. 
By including the full face region, the models become more susceptible to such interference from hand movements near the mouth area. 
Given the complementary nature of face and lip ROI images under varying conditions,
it is intuitive to combine both the face-based and the lip-based models
for jointly recognizing the speech content.
The model ensemble strategy was employed, treating either the lip-based or the face-based models as base models.

The experimental results are shown in Table~\ref{tab:tab5}. 
For comparison, each model was also ensembled with its counterpart trained with a different random seed, denoted as M1+M1*, M2+M2*, M3+M3*, and M4+M4*, respectively. 
From Table~\ref{tab:tab5}, ensembling models
with different random seeds achieved performance
comparable to the corresponding single models.
However, when ensembling the face-based and lip-based models, 
a substantial improvement over the single models 
 was attained. Specifically, M1+M2 achieved a relative CER reduction of 3.1\% 
compared to the average CER of M1 and M2.
M3+M4 yield a 3.8\% CER reduction over the average CER of
M3 and M4.
Eventually, when ensembling all single models, M1+M2+M3+M4 obtained
the lowest lipreading CER of 65.8\% on our ChatCLR validation set.

Figure~\ref{fig:figure7} presents illustrative examples of lipreading results along with their corresponding face and lip ROI video frames. In the top two cases of the figure, the lip-based model outperformed the face-based model, possibly because the speaker's head holding a microphone near the mouth may have more significantly distracted the face-based model. In contrast, in the bottom two cases of the figure, the face-based model achieved fewer recognition errors. 
Additionally, it can be observed that our proposed model ensemble strategy, which aggregates predictions from both the lip ROI-based and face-based models, effectively enhances recognition performance.

\begin{table}
    \centering
    \begin{tabular}{cccc}
    \hline
    \hline
       Model ID & Method & ESB &  CER (\%)  $\downarrow$  \\
       \hline
        M1 & AV2vec-tf-lip & \multirow{4}{*}{\XSolidBrush} & 77.2 $\pm$ 1.6 \\
        M2 & AV2vec-tf-face & & 74.6 $\pm$ 1.5 \\
        M3 & AV2vec-lip  &    & 72.6 $\pm$ 1.6 \\
        M4 & AV2vec-face &    & 68.4 $\pm$ 1.7 \\
        \hline
        M1+M1* & -- & \multirow{7}{*}{\Checkmark} & 77.6 $\pm$ 1.5 \\
        M2+M2* & -- &  & 74.0 $\pm$ 1.5 \\
        M3+M3* & -- &  & 72.9 $\pm$ 1.6 \\
        M4+M4* & -- & & 68.2 $\pm$ 1.7 \\
        M1+M2 & -- &  & 72.8 $\pm$ 1.7 \\
        M3+M4 & -- &  & 66.7 $\pm$ 1.8 \\
        M1+M2+M3+M4 & -- & & 65.8 $\pm$ 1.8 \\
        \hline
        \hline
    \end{tabular}
    \caption{Lipreading results of speaker-dependent (SD) single model and ensembled models on ChatCLR  validation set.  $\pm$ indicates 95\% confidence intervals. ``*'' represents the model was trained with the 
    same configuration but a different random seed. ``ESB'' represents model ensemble.}
    \label{tab:tab5}
\end{table}

\subsubsection{Results on the 2024 ChatCLR Challenge}
\label{sec:subsubChall}

\begin{table}
    \centering
    \scalebox{0.9}{
    \begin{tabular}{lc}
    \hline
    \hline
        \multicolumn{2}{c}{Challenge leaderboard} \\
        \hline
            Team &   CER (\%)  $\downarrow$  \\
            \hline
        ChatCLR-baseline &  99.7 \\
       \textbf{snnu-lip (Ours, M1+M2+M3)}  &     78.7 \\
       JaeEunBaik$^\dagger$ &   78.4 \\
       OneTakeLove~\citep{10645400} &   	78.2 \\
       SparkFly~\citep{10645443}  &   77.6 \\
       \hline
        \multicolumn{2}{c}{Our method} \\
        \hline
        Model &   CER (\%)  $\downarrow$  \\
        \hline
    M1: AV2vec-tf-lip  & 84.4 $\pm$ 1.2 \\
    M2: AV2vec-tf-face  & 83.9 $\pm$ 1.2 \\
    M3: AV2vec-lip  &  82.0 $\pm$ 1.3 \\
    M4: AV2vec-face  & 81.4 $\pm$ 1.3 \\
    M1+M2             &  80.9   $\pm$ 1.3 \\
    M3+M4             &  78.7  $\pm$ 1.5 \\
    \textbf{M1+M2+M3+M4}       &  \textbf{77.3} $\pm$ 1.5 \\
    \hline 
    \hline
    \end{tabular}}
    \caption{ Results of the baseline, top 4 ranked teams  results in the task 2 of 2024 ChatCLR
    Challenge, and our proposed method on the official ChatCLR evaluation set. $\pm$ indicates 95\% confidence intervals. $^\dagger$The team's method remains unpublished.
    The single models M1 to M4 are all speaker-dependent.
    %``+'' represents model ensemble. 
   The training of M4 
   was unfinished by the end of the Challenge,
   hence only M1 to M3 were ensembled for the
   competition.}
    \label{tab:tab6}
\end{table}

During the competition, our models were not fully trained, and only three models, namely AV2vec-tf-lip, AV2vec-tf-face, and AV2vec-lip, were ensembled. This ensemble obtained the fourth place out of seven teams in task 2 of the 2024 ChatCLR Challenge. Subsequently, with the assistance of the Challenge organizers, we were able to evaluate our final results on the official ChatCLR evaluation set. The results are summarized in Table 6. As evident from the table, our complete method, with an ensemble of 4 base models, achieved a lower CER than the top-ranked team in the challenge.
However, statistical analysis at the 95\% confidence level revealed no significant differences 
between our approach and the top-performing teams. 
%Nonetheless, these competitive results suggest that our method performs on par with state-of-the-art approaches.
 Future work will aim for statistically significant improvements over existing methods. 

\section{Discussion}
Although our method has achieved competitive lipreading result on the evaluation set of the ChatCLR corpus,
the character error rate of 77.3\% remains high, indicating the inherent difficulty of this task. Consequently, there is still considerable room for improvement before lipreading becomes truly practical.
It is worth noting that a single AV2vec model achieves a word error rate (WER) of 34.3\% on the LRS3 test set, significantly lower than 
the 77.3\% CER of our method on the ChatCLR, although
they cannot be directly compared.
This stark discrepancy underscores that the lipreading model is highly susceptible to variations in language, speaking styles and data quality.

One crucial way to further enhance
lipreading performance for this task
is to collect or  synthesize more audio-visual data in Chinese.
Transfer learning from a larger amount of videos in other 
language can also be an effective method, as 
demonstrated in our experiments.
Moreover, our work
has demonstrated the complementary
nature of the lip ROI and the face as the visual input for lipreading
in realistic and complex situations. An ensemble 
learning method is proposed to combine both types of inputs.
However, a more computationally efficient
method can possibly be designed to  dynamically select
inputs for lipreading, maximizing performance.
Nevertheless, the model ensemble still serves as
an intriguing and well-performed baseline method for utilizing diverse regions in the lipreading task.

%In this work, we focus on using full parameters finetuning during speaker adaptation.
%Recently, parameter-efficient finetuning (PEFT) has emerged as 
%an important techniques for adaptation of large pretrained models.
%In our experiments, the base and large version of the AV2vec model have about 100 M and 300 M trainable parameters,
%therefore unfreezing full parameters during adaptation is still an affordable way.
%Compared to finetuning, pretraining phase costs most our time as described in Section~\ref{sec:modelconfig}.
%A recent work~\cite{gimeno2023comparing} indicates the Adapter-based method
%achieved dramatic time reduction during finetuning but sacrificed the lipreading recognition accuracy.
%We also plan to further explore the PEFT-based method for lipreading speaker adaptation in our future work.

In this work, we focus on full-parameter finetuning during speaker adaptation. 
While parameter-efficient finetuning (PEFT) has emerged as a significant technique for adapting large pretrained models, 
our models contain approximately 170M trainable parameters, 
making full-parameter unfreezing during adaptation still computationally feasible. 
Compared to the finetuning phase, pretraining consumed the majority of our computational resources, 
as detailed in Section~\ref{sec:modelconfig}. 
Recent research~\citep{gimeno2023comparing} demonstrates that while Adapter-based methods achieve substantial time reduction during fine-tuning, 
they yield suboptimal lipreading recognition accuracy. 
We plan to further investigate PEFT-based methods for lipreading speaker adaptation in our future work.

%We will leave this for our future studies.

\section{Conclusion}

In this work, we extend our previously proposed AV2vec by introducing
a cross-lingual transfer learning strategy for visual representation learning.
To improve the lipreading accuracy for the target speaker,
speaker adaptation with KL divergence regularization is employed to
obtain speaker-dependent models. For the 
input of the lipreading model, both conventional lip region-of-interest (ROI) and the entire face
are compared. During testing, a model ensemble strategy
is
proposed to aggregate 
predictions from both lip and face-based models, enhancing
the robustness and the performance in the lipreading task.
Our experimental results demonstrated that the cross-lingual transfer learning
 not only saved training costs but also leaded to 
better lipreading results  when the data in the target language is limited.
Our experiments also  validated the effectiveness 
of the speaker adaptation and the proposed strategy of ensembling models with
the lip ROI and the face input.
Eventually, our proposed method achieved a lower character error rate 
 than the top-ranked team in the
2024 ChatCLR Challenge on the ChatCLR evaluation set.

\section{Acknowledgements}

This work was supported by the
 Fundamental Research Funds for the Central Universities (Grant No. GK202406005),
National Natural Science Foundation of China  (Grant No. 62401348, 62077035, 61977044, 62206162), the Ministry of Education’s Cooperative Education Project (Grant No. 202102591018),
and Young Talent Fund of Xi'an Association for Science and Technology (Grant No. 959202313048).
The authors
gratefully acknowledge 
the support from
the organizers of the 2024 ChatCLR Challenge  for their 
assistance in evaluating our results
after the competition.

\section{Declaration of Generative AI and AI-assisted technologies in the writing process}

During the preparation of this work, the authors used iFLYTEK's Xinghuo\footnote{https://xinghuo.xfyun.cn/} in order to improve readability and language. After using this tool, the authors reviewed and edited the content as needed and take full responsibility for the content of the publication.

\begin{table}[h]
    \centering
    \scalebox{0.75}{
    \begin{tabular}{cc}
    \hline
    \hline
    \multicolumn{2}{c}{\textbf{Audio frontend}} \\
    \hline
    stage  &  Input log filter bank sequence ($T \times 104 $) \\
    $\text{fc}_{1}$      &  fully connected, 512 \\
    \hline
    \multicolumn{2}{c }{ \textbf{Video frontend}} \\
    \hline
      stage   &  Input image sequence $(T \times 88 \times 88)$ \\
      \multirow{2}{*}{$\text{conv}_1$}   & conv3d, $5 \times 7 \times 7 $, 64, stride $1\times 2 \times 2 $ \\
         &  maxpool, $1 \times 3 \times 3$\\
       $\text{res}_2$  &   $\begin{bmatrix} \text{conv2d}, 3 \times 3, 64 \\ \text{conv2d}, 3 \times 3, 64 \end{bmatrix} \times 2 $ \\
        $\text{res}_3$  & $\begin{bmatrix} \text{conv2d}, 3 \times 3, 128 \\ \text{conv2d}, 3 \times 3, 128 \end{bmatrix} \times 2 $ \\
        $\text{res}_4$  & $\begin{bmatrix} \text{conv2d}, 3 \times 3, 256 \\ \text{conv2d}, 3 \times 3, 256 \end{bmatrix} \times 2 $ \\
        $\text{res}_5$  & $\begin{bmatrix} \text{conv2d}, 3 \times 3, 512 \\ \text{conv2d}, 3 \times 3, 512 \end{bmatrix} \times 2 $ \\
        $\text{pool}_6$ & global average pooling \\
    \hline
    \multicolumn{2}{c}{\textbf{Modality fusion}} \\
    \hline
    stage & Input representations $(T \times 1024)$ \\
    $\text{fc}_1$   & layer norm, fully connected, 768 \\
    \hline
    \multicolumn{2}{c}{\textbf{Transformer encoder blocks}} \\
    \hline
    stage & Input representations $(T \times 768)$ \\
    $\text{pos}_1$ & conv1d, 128, group 16 \\
    $\text{trans}_2$ & $ \begin{bmatrix}
        \text{layer norm,} 768\\
        \text{self attention}, \text{head 12} , 768 \\
        \text{layer norm,} 768 \\
        \text{ffn}, 768 \rightarrow 3072 \rightarrow 768 \\
    \end{bmatrix} \times 12 
    $ \\
    $\text{ln}_3$ & layer norm, 768 \\
    \hline
        \multicolumn{2}{c}{\textbf{Transformer decoder blocks}} \\
        \hline
    stage & Input text sequence (N) \\
    $\text{emb}_1$ & word embedding, $6000 \times 768$ \\
    $\text{trans}_2$ &  $ \begin{bmatrix}
        \text{layer norm,} 768\\
        \text{self attention}, \text{head 12} , 768 \\
         \text{layer norm,} 768\\
        \text{cross attention}, \text{head 12} , 768 \\
        \text{layer norm,} 768 \\
        \text{ffn}, 768 \rightarrow 3072 \rightarrow 768 \\

    \end{bmatrix} \times 6 
    $ \\
        $\text{ln}_3$ & layer norm, 768 \\
        $\text{out}_4$ & fully connected, 6000 \\
    \hline  
    \hline
    \end{tabular}
    }
    \caption{The details of model architecture for our proposed method. 
    Audio frontend was only adopted during AV2vec pretraining.
    Decoder was adopted except AV2vec pretraining. Residual connection is omitted in this table.}
    \label{tab:table2}
\end{table}

\appendix
\section{Details of model architecture}\label{app1}

The details of our model architecture are presented in Table~\ref{tab:table2}.

\bibliographystyle{elsarticle-harv} 
\bibliography{ref}

\end{document}